\documentclass[useAMS,usenatbib]{mn2e}
\usepackage{epsfig}
\usepackage{times}

\newcommand{\target}{Swift~J1753.5--0127}
\newcommand{\xte}{\textit{RXTE}}
\newcommand{\rxte}{\textit{RXTE}}

\title[Echo Mapping of \target]{Echo Mapping of \target}
\author[R. I. Hynes et al.]{R. I. Hynes$^{1}$\thanks{E-mail:
rih@phys.lsu.edu (RIH)}, K. O. Brien$^{2}$, F. Mullally$^{3}$ and T. Ashcraft$^{4}$\\
$^{1}$Department of Physics and Astronomy, Louisiana State
University, Baton Rouge, Louisiana 70803, USA\\
$^{2}$European Southern Observatory, Casilla 19001,
        Santiago 19, Chile\\
$^{3}$Department of Astrophysical Sciences, Princeton University, Princeton, NJ, 08544, USA\\
$^{4}$School of Earth and Space Exploration, Arizona State University, Tempe, AZ 85287-1404, USA}
\begin{document}

\date{Accepted ??. Received ??; in original form ??}

\pagerange{\pageref{firstpage}--\pageref{lastpage}} \pubyear{2009}

\maketitle

\label{firstpage}

\begin{abstract}

  We present two epochs of coordinated X-ray--optical timing
  observations of the black hole candidate \target\ during its 2005
  outburst.  The first epoch in July occurred at outburst peak.  Two
  consecutive nights of observations using the McDonald Observatory
  Argos camera with the Rossi X-ray Timing Explorer show a consistent
  correlation with an immediate response and an extended tail lasting
  $\sim5$\,s.  The properties of the variability and the correlation
  are consistent with thermal reprocessing in an accretion disk.  The
  shortness of the lag suggests a short orbital period consistent with
  that recently claimed.  The second epoch in August used the VLT
  FORS2 HIT mode again in conjunction with \xte. Again a repeatable
  correlation is seen between two independent subsets of the data.  In
  this case, though, the cross-correlation function has an unusual
  structure comprising a dip followed by a double-peak.  We suggest
  that this may be equivalent to the dip plus single peak structure
  seen by \citet{Kanbach:2001a} in XTE~J1118+480 and attributed there
  to synchrotron emission; a similar structure was seen during later
  activity of \target\ by \citet{Durant:2008a}.

\end{abstract}


\section{Introduction}

The X-ray source \target\ was the first transient black hole candidate
discovered by the {\it Swift} mission on 2005 June 30
\citep{Palmer:2005a,Burrows:2005a}.  It was also detected at high
energies by \rxte\ \citep{Morgan:2005a} and {\it INTEGRAL}
\citep{CadolleBel:2005a}. The X-ray source was associated with an
optical counterpart with $R=15.8$ \citep{Halpern:2005a} and a blue
spectrum showing H$\alpha$ and He\,{\sc ii} emission
\citep{Torres:2005a,Torres:2005b}.  A variable but unresolved radio
counterpart was also identified by \citet{Fender:2005a}.

From the outset of the outburst, X-ray and $\gamma$-ray observations
indicated that the spectrum and variability properties of \target\
were typical of a black hole system in the hard state
\citep{Morgan:2005a,CadolleBel:2005a}.  The presence of a compact
radio component is also consistent with this \citep{Fender:2005a}.
While canonical black hole systems often enter a soft thermally
dominated state near outburst peak, there are a number of other
systems that have remained in the hard state throughout
\citep{Brocksopp:2004a}.

Relatively little is known about the binary system.  We identified
significant optical variability and short timescale correlations with
X-rays \citep{Hynes:2005a,Hynes:2005b}.  Interpreting the small lags
seen as due to thermal reprocessing in an accretion disk around a
black hole we suggested a relatively short period of less than 12\,hrs
\citep{Hynes:2006a}.  This was subsequently supported by optical
photometric observations that found a 3.24\,hr periodicity interpreted
as a superhump modulation \citep{Zurita:2007a,Zurita:2008a}.  If this
interpretation is confirmed then the orbital period will be slightly
less than this making this the shortest orbital period black hole
candidate yet discovered.

In this work we present two epochs of coordinated X-ray and optical
observations at high time-resolution during the main outburst in 2005,
one in early July and the other in early August.  Preliminary results
from this study were presented by \citet{Hynes:2005a},
\citet{Hynes:2005b}, \citet{Hynes:2006a} and \citet{Hynes:2007a}.  We
note that \citet{Durant:2008a} reported similar, but not exactly the
same behavior as we see in our August observation during a later phase
of activity in 2007.

\section{Observations}

\subsection{SMARTS 1.3\,m Andicam data}

We performed daily monitoring, where possible, using the SMARTS 1.3\,m
telescope at Cerro Tololo, Chile, for the first three months of the
outburst until visibility became compromised.  Observations used the
Andicam dual-channel optical/IR camera.  Two 130\,s $V$ band
observations were performed together with five $50$\,s dithered $H$
band images each night.

All data reduction used standard procedures in {\sc iraf}.  Optical
data were supplied with satisfactory pipeline reduction.  IR data from
each night were median filtered to create a sky image which was then
subtracted from each frame before shifting and co-adding the
individual images.  In both bands aperture photometry was performed
relative to a comparison star in the field.  

The optical comparison star was calibrated relative to several
standard stars in the field of T~Phe \citep{Landolt:1992a}.  The IR
comparison star was calibrated relative to 2MASS photometry.  In both
cases the uncertainty in the calibration was $\sim0.04$\,mag.

Our $V$ and $H$ lightcurves are shown in Fig.~\ref{ASMFig}.  We have
supplemented these lightcurves with additional points from
\citet{Still:2005a}, \citet{Torres:2005a}, and \citet{CadolleBel:2007a}.

\subsection{McDonald 2.1\,m Argos data}

Fast optical photometry was obtained on 2005 July 6 and 7 for
approximately 1.5\,hrs each night using the Argos CCD photometer on
the McDonald Observatory 2.1\,m telescope \citep{Nather:2004a}.
Observations on each night were performed as an uninterrupted
time-series of 1\,s exposures using a $V$ filter; there was no
dead-time between exposures.  Absolute timing was synced to GPS 1\,s
ticks and multiple NTP servers.

Custom IDL software written for the instrument was used to subtract
dark current and bias and flat-field the data before performing
aperture photometry of \target\ and a brighter comparison star using a
2\,arcsec radius aperture.

\subsection{VLT 8\,m FORS2/HIT data}
\label{FORSSection}

Further rapid observations were performed on 2005 August 9 using the
HIT fast imaging mode \citep{OBrien:2007a} of the FORS2 instrument on
the European Southern Observatory's UT2/VLT telescope on Cerro
Paranal, Chile. The observation were taken with the V-band Bessel
filter. The HIT mode uses a slit to allow only a small fraction of the
CCD to be exposed to light. It then shifts the charge from this
exposed region at regular intervals and stores it on the un-exposed
regions of the CCD. Once the masked region is full (after accumulating
data for 64\,s), the shutter is closed and the CCD read out in the
usual manner. The slit-width used in the observations was 3\arcsec,
which projects onto 12 (binned) pixels implying a time-resolution of
0.5\,s. The image was very stable, with an rms deviation of
0.14~pixels during the entire run. The image quality was equally
stable with a mean FWHM of 2.44~pixels (0.61\arcsec) and an rms of
0.2~pixels. The flux was extracted using a variable 1-D aperture along
the slit on the bias subtracted images. The width of the aperture was
set to be 5$\times$FWHM, as measured from a fit to the comparison star
profile. The sky was measured in symmetric regions around this. An
identical aperture was used to extract the flux from the comparison
star.

\subsection{\xte/PCA data}

All high time-resolution observations were coordinated with public
\xte/PCA observations.  For this analysis we used Standard 1 mode
lightcurves with 125\,ms time-resolution and all energy channels
merged into a single bin.  Each lightcurve was extracted from final
data products using standard tools.

\section{The Outburst Lightcurve}

\begin{figure}
\epsfig{width=3.2in, file=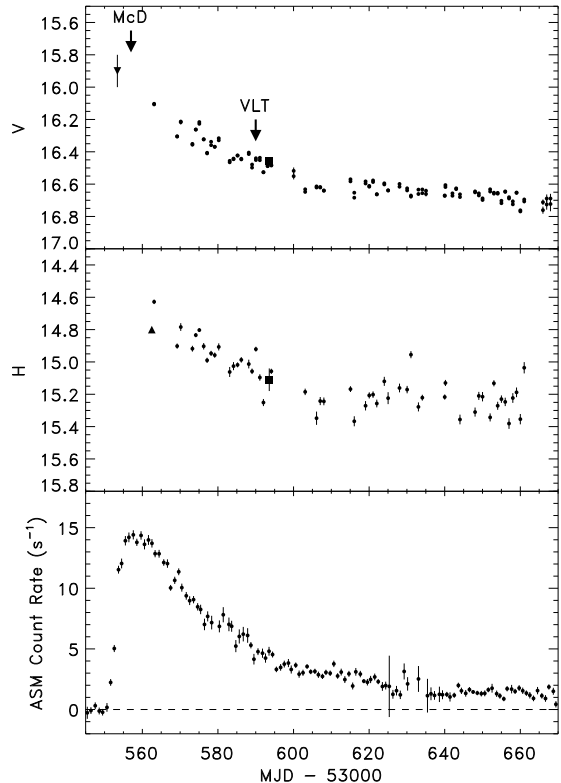}
\caption{Lightcurves of the early outburst.  The upper two panels are
  from SMARTS data, the lower from \xte/ASM
  monitoring.  The downward pointing
  triangle shows the Swift/UVOT measurement of Still et al.\ (2005).
  The upward pointing triangle is from Torres et al.\ (2005), and
  square points are from Cadolle Bel et al.\ (2007).  Times of
  coordinated observations are shown.}
\label{ASMFig}
\end{figure}

In Fig.~\ref{ASMFig} we show optical, IR, and X-ray lightcurves
spanning the first three months of outburst.  The source followed a
fast-rise, slow decay pattern with a plateau seen around
60\,days after the outburst peak.  This is a very common morphology
for a transient black hole outburst \citep{Chen:1997a}.  We also
indicate the times of our rapid optical observations.  The first
occurred almost exactly at the outburst peak, the second about a month
later on the decline, somewhat before the plateau.  The
optical observations of \citet{Zurita:2008a} and \citet{Durant:2008a}
were obtained much later, beyond the coverage here.

\section{Lightcurves}

\subsection{Outburst Peak}

\begin{figure*}
\epsfig{angle=90,width=6.4in, file=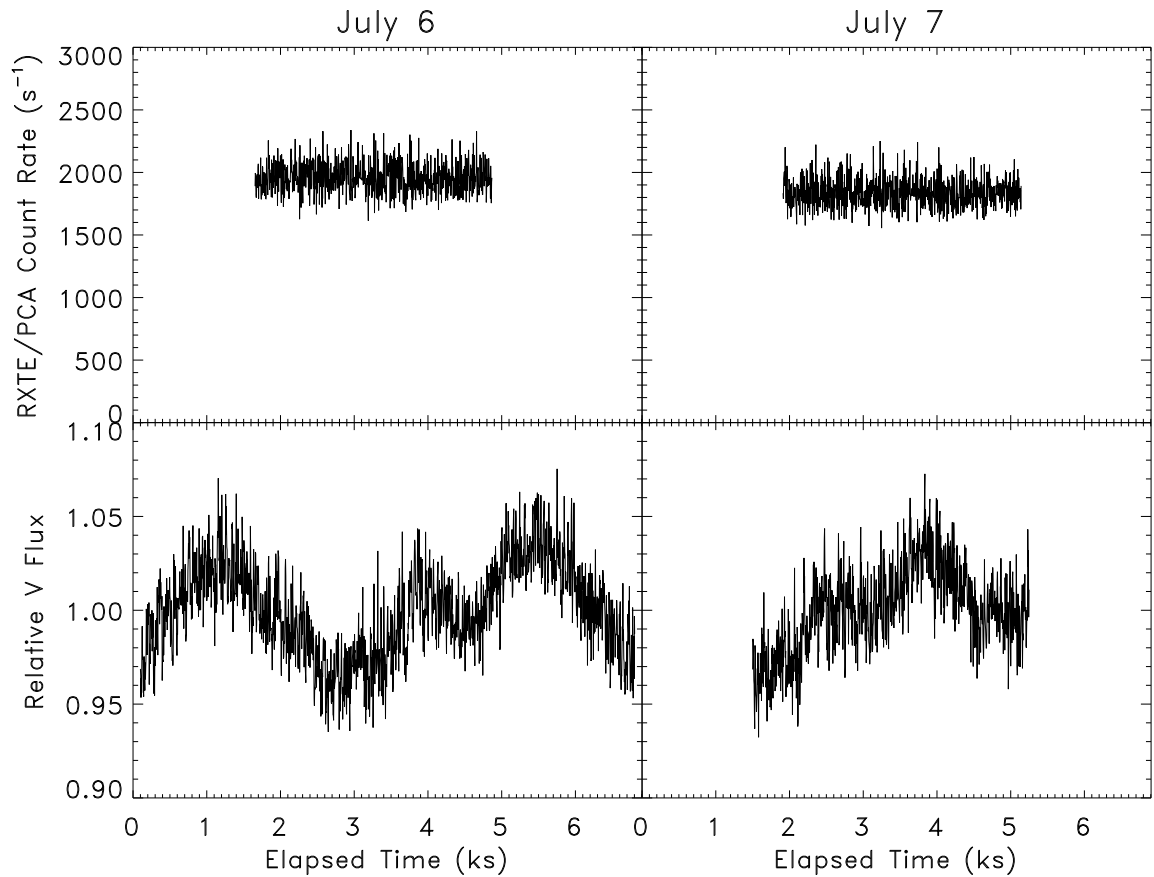}
\caption{Lightcurves from our first coordinated observations.  All
  have been binned to 4\,s time-resolution for clarity and an arbitrary
  offset in time has been applied (while preserving the relative
  timing of the X-ray and optical lightcurves.)}
\label{ArgosLCFig}
\end{figure*}

The lightcurves from July 6 and 7, near the outburst peak,
are shown in Fig.~\ref{ArgosLCFig}.  The X-ray lightcurves show no
long term variations, but considerable rapid variability
(r.m.s. $\sim12$\,percent on timescales longer than 1\,s compared to a
statistical uncertainty of 2\,percent).  The optical data, on the
other hand, show substantial lower-frequency variability.  The data
appear consistent with a 1.2\,hr period, although they are
insufficient to prove that these observations are truly periodic
rather than stochastic.  If real, this periodicity does not correspond
to either the proposed orbital period $\sim3.2$\,hr
\citep{Zurita:2008a} or the first harmonic of it.  Spurious apparent
periods are not unusual in outburst (e.g.\ \citealt{Leibowitz:1991a}) so
we do not believe the 1.2\,hr `period' is real and disregard it.

\subsection{Outburst Decline}

\begin{figure}
\epsfig{angle=90,width=3.2in,file=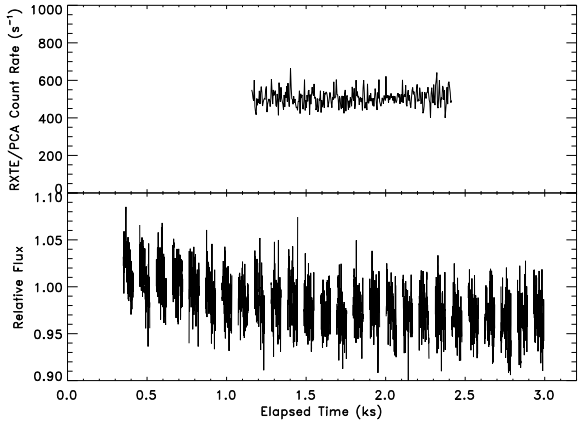}
\caption{Lightcurves from our second coordinated observation.  All
  have been binned to 4\,s time-resolution for clarity and an arbitrary
  offset in time has been applied (while preserving the relative
  timing of the X-ray and optical lightcurves.)}
\label{VLTLCFig}
\end{figure}

The lightcurves from August 9 are shown in Fig.~\ref{VLTLCFig}.  The
source is clearly considerably fainter in X-rays, but considerable
variability is still present (r.m.s. $\sim15$\,percent on timescales
longer than 1\,s).  It is hard to directly compare these with the July
optical observations due to the very different characteristics of the
data.  Both slow and rapid optical variations do appear to be present,
however, as in the previous datasets.

\section{Power Density Spectra}

\begin{figure}
\epsfig{angle=90,width=3.2in,file=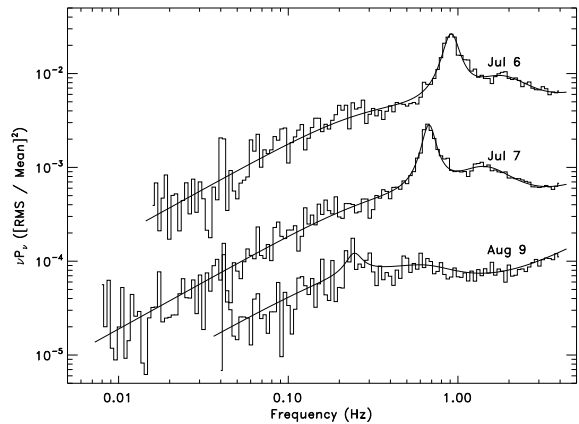}
\caption{Power density spectra of \xte\ data from the three
  simultaneous observations.  The fits, each of three Lorentzians plus
  white noise, are described in the text.  The July 7 PDS has been
  scaled down by a factor of 10 and the August 9 one by a factor 100
  for clarity.}
\label{PDSFig}
\end{figure}

To better characterise the temporal properties at each epoch we have
calculated power density spectra (PDS) of each \xte/PCA lightcurve
and show them in Fig.~\ref{PDSFig}.  The general form of each PDS is
band-limited noise, resembling a 
broken power-law, characteristic of the hard state.  The data
from July additionally includes a strong low-frequency
quasi-periodic oscillation close to 1\,Hz.  We fit these PDS with a
multi-Lorentzian decomposition following \citet{Belloni:2002a}.  The
band-limited noise is represented by a zero-centred Lorentzian (i.e.\
one with zero centroid frequency) and the QPO by a peaked one (i.e.\
one with non-zero centroid frequency).  Even when white noise is
included, we find pronounced residuals at a few Hz, so include a
second broader QPO.  This is a similar model to that adopted by
\citet{CadolleBel:2007a} except that we allow the third Lorentzian
to be a peaked QPO rather than a band-limited noise component.

\begin{table}
  \caption{Characteristic frequency ($\nu_{\rm max}$) and amplitude ($r$) of multi-Lorentzian fits to X-ray PDS.}
\label{PDSFitTable}
\begin{tabular}{llcc}
\hline
\noalign{\smallskip}
Date  & Component & $\nu_{\rm max}$ (Hz) & $r$ \\
\noalign{\smallskip}
\hline
\noalign{\smallskip}
Jul 6 & BLN       & $0.46\pm0.04$ & $0.136\pm0.004$ \\
      & QPO       & $0.940\pm0.014$  & $0.098\pm0.006$ \\ 
      & QPO2      & $2.5\pm0.2$    & $0.10\pm0.02$   \\ 
\noalign{\smallskip}
Jul 7 & BLN       & $0.67\pm0.06$ & $0.141\pm0.005$ \\
      & QPO       & $0.685\pm0.005$  & $0.089\pm0.003$ \\ 
      & QPO2      & $1.88\pm0.06$    & $0.106\pm0.006$   \\ 
\noalign{\smallskip}
Aug 9 & BLN       & $0.64\pm0.12$ & $0.20\pm0.02$ \\
      & QPO       & $0.251\pm0.014$  & $0.045\pm0.010$ \\       
      & QPO2      & $0.87\pm0.13$    & $0.06\pm0.02$   \\ 
\noalign{\smallskip}
\hline
\end{tabular}
\end{table}

With these components, good fits can be obtained for all three nights;
we summarize the characteristic frequency and amplitude of each
component in Table.~\ref{PDSFitTable}.  The QPOs were less pronounced
on August 9 than they were in July but with lower amplitudes the same
model provides a good fit, and a QPO does still appear to be required.
We did have to fix the widths of the QPOs on August 9 to the average
values of the July observations to adequately constrain the fit.  The
most important quantity for our later correlation analysis is the
characteristic period of the primary QPO.  This is 1.06\,s for July 6,
1.46\,s for July 7, and 4.0\,s for August 9.  Since the Argos
time-resolution is only 1\,s, we would not expect the QPOs to be
resolved in the July optical data and their main effect is likely to
be indistinguishable from an excess of white noise.  The 4.0\,s QPO in
August could have a much more significant effect on our analysis.

\section{Cross Correlation Functions}

\subsection{Outburst Peak}

\begin{figure}
\epsfig{width=3.2in,file=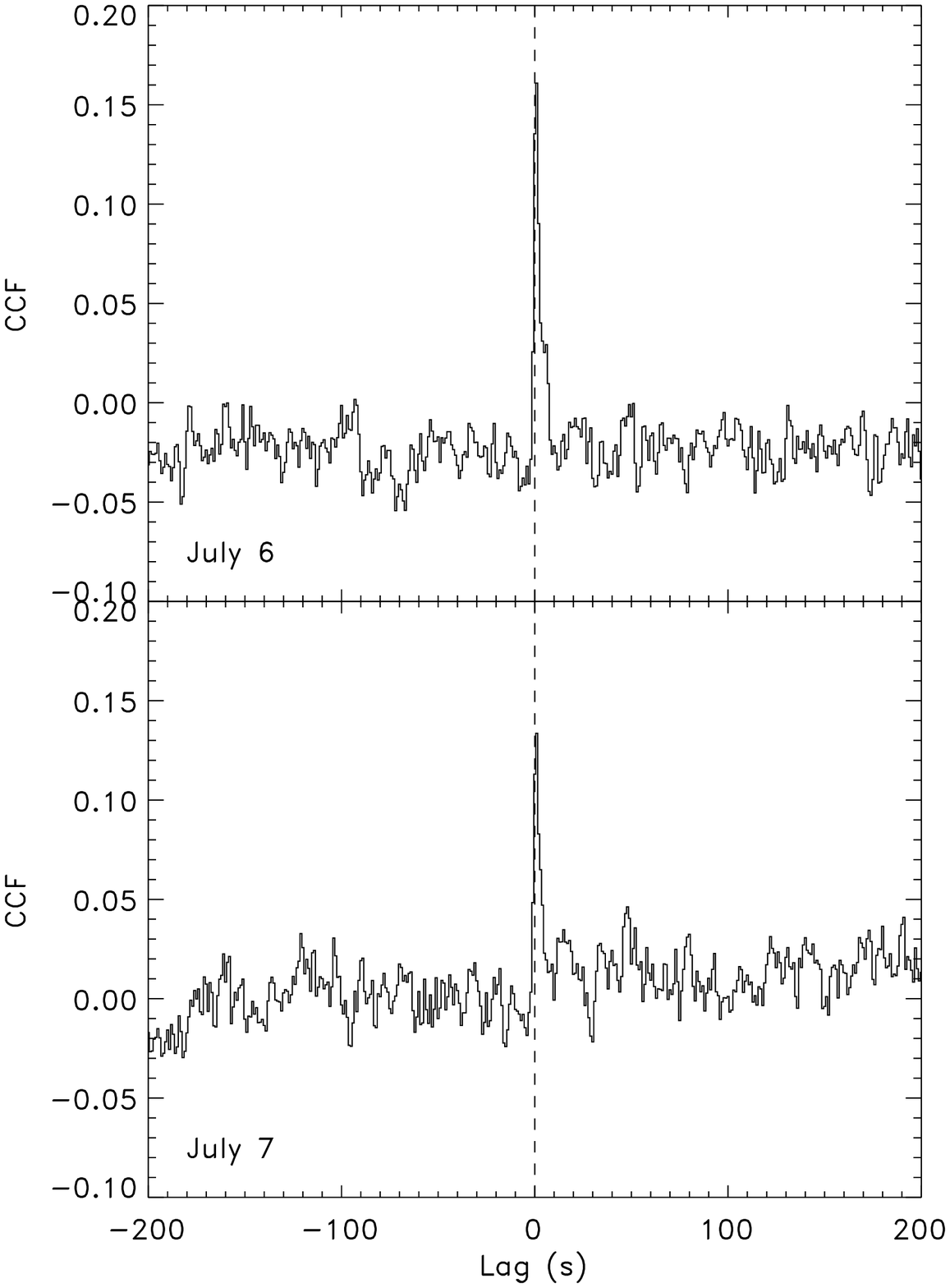}
\caption{Cross correlation functions for July data from the outburst peak.  Positive lags indicate that the optical lags behind the X-rays.  No detrending or filtering has been done to the data.}
\label{ArgosCCFFig}
\end{figure}

Cross correlation functions (CCFs) for the July 6 and 7 observations near
the outburst peak are shown in Fig.~\ref{ArgosCCFFig}. Although
correlations were not obvious in the lightcurve they are clearly
present at high significance, with a strong response at or close to
zero lag, and an extended tail lasting for a few seconds.

The obvious interpretation of the correlation is that it arises from
thermal reprocessing either in an accretion disk or on the surface of
the companion star.  The lag of the companion star response modulates
with orbital phase with an average of approximately the binary
separation in lightseconds.  With the parameters discussed in
Section~\ref{TransferFunctionSection}, the binary separation would be
about 5\,lightseconds and so the companion peak lag would modulate in
the 0--10\,s range.  Since our observations span 1.5\,hrs each night
and the orbital period is estimated to be 3.2\,hrs
\citep{Zurita:2008a}, we cover about half an orbit on each night.  On
both nights we observe a peak lag close to zero, so it is unlikely
that the companion is dominating the response, which instead appears
to originate from the accretion disk.  Additionally, black hole
systems are expected to usually have small mass ratios, and the
presence of likely superhumps \citep{Zurita:2007a} supports this
expectation in this case.  If the mass ratio is indeed small, then the
companion star will subtend only a small solid angle as seen by the
X-ray source, and hence would be expected to reprocess much less light
than the accretion disk.

We should be somewhat cautious, however, as similar CCFs were seen in
the UV in XTE~J1118+480 (although not in the optical), where they may
be associated with jet emission.  This seems less likely in this case,
however, as the correlations are observed at the peak of an apparently
normal outburst, and the optical spectrum is blue and can be fitted by
that of an accretion disk \citep{Still:2005a}.  In addition, the
optical auto-correlation function is found to be broader than the
X-ray one as expected for thermal reprocessing, and unlike that seen in
XTE~J1118+480.  We cannot conclusively rule this possibility out,
however, especially as the CCFs seen in August (at
lower luminosity) do resemble those of XTE~J1118+480.

\subsection{Outburst Decline}

\begin{figure}
\epsfig{angle=0,width=3.2in,file=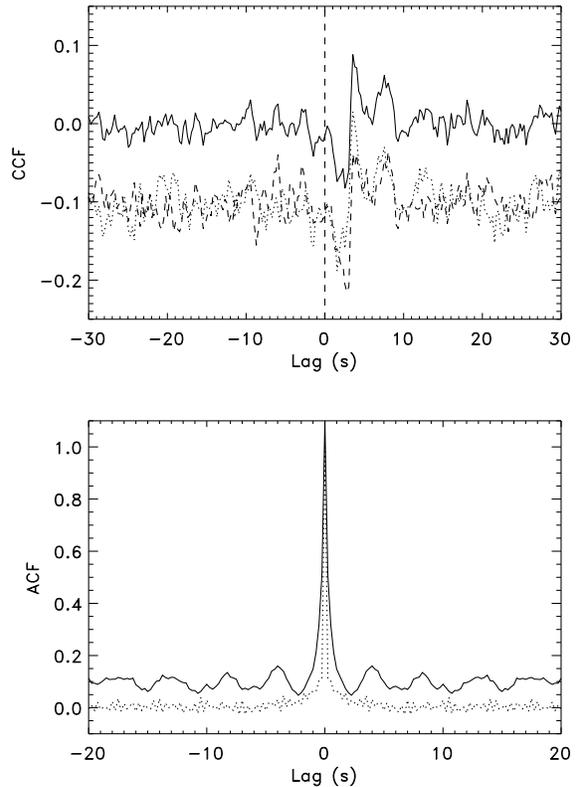}
\caption{Cross-correlation and auto-correlation functions for August 9
  data from the outburst decline.  In the upper panel CCFs, the upper
  solid line is computed from the full dataset, the lower dashed and
  dotted lines separate odd and even optical observations to
  demonstrate the repeatability of the structures seen.  In the lower
  ACFs, the solid curve is the X-ray ACF which has been offset by
  $+0.1$ and the dotted curve is the optical ACF.  The latter includes
  a large component of statistical noise at lag 0, and may also be
  affected at small lags by the blurring of the data in the time-axis
  introduced by the data acquisition method (see
  Section~\ref{FORSSection}).}
\label{VLTCCFFig}
\end{figure}

We have similarly calculated CCFs for August 9 
using VLT data.  We do not then have two independent
observations, however we can compute independent CCFs by using odd and
even segments of the optical data.  For these data we computed CCFs
for each optical segment separately and co-added them.  We show both
the odd and even CCFs, and the combined CCF in
Fig.~\ref{VLTCCFFig}.

The results are somewhat puzzling.  A correlation is clearly present,
with the structure of the feature well reproduced in both independent
CCFs.  The two observatories are obviously themselves completely
independent, so instrumental effects cannot be responsible and this
feature must be intrinsic to the source.  The shape of the
CCF is strange, exhibiting a dip followed by two pronounced peaks.
This general structure is not unprecedented.  A similar dip followed
by peak structure was seen by \citet{Kanbach:2001a} in XTE~J1118+480,
and later in the activity of \target\ \citet{Durant:2008a} found a
strong dip followed by a weak peak.

The overall dip plus peak structure, and indeed the timescales
involved, are rather similar to those seen in XTE~J1118+480, and less
so to the later observations of \target.  The double peak has not been
seen in other sources.  One possible explanation is that this is
actually related to the weak QPO seen at this epoch.  We deduced a
4.0\,s periodicity for this, and the two main CCF peaks in
Fig.~\ref{VLTCCFFig} are separated by 4.0\,s.  We therefore interpret
the splitting of the peak as a result of the QPO aliasing the CCF
structure.

\section{Model Transfer Functions}
\label{TransferFunctionSection}

If we interpret the July correlations as arising from thermal
reprocessing we can investigate them further by comparing them with
predicted disk transfer functions from \citet{OBrien:2002a}.  We are
of course at a disadvantage in not having a full set of system
parameters.  As a starting point, we can however make plausible
assumptions.  The most reasonable is that the orbital period is indeed
close to 3.2\,hr as suggested by \citet{Zurita:2008a}.  Others are
largely guesses.  We assume a representative black hole mass of
7\,M$_{\odot}$, a mass ratio of 0.1 (reasonable for a short-period
system believed to exhibit superhumps), an inclination of 60$^{\circ}$
(the most probable inclination to be drawn from a random
distribution), and a tidally truncated disk.

\begin{figure}
\epsfig{width=3.2in,file=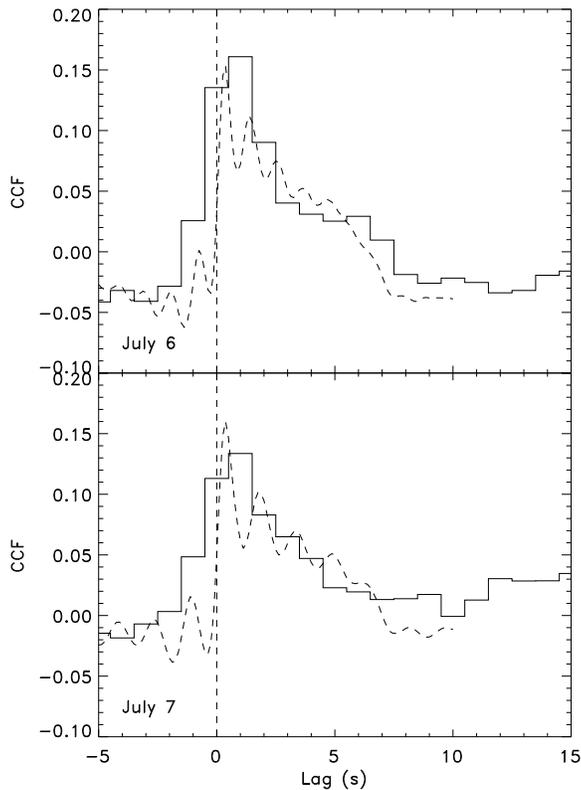}
\caption{Cross correlation functions for July data from the outburst
  peak (shown as histograms) compared with those expected for disk
  transfer functions (shown dashed).}
\label{ModelCCFFig}
\end{figure}

We calculate the transfer function following the methods of
\citet{OBrien:2002a}.  As done by \citet{Hynes:2003a} we then convolve
the transfer function with the X-ray auto-correlation function to
obtain a predicted cross-correlation function.  This is more useful
than attempting to compare with the lightcurves themselves when
correlated features cannot be directly seen in the lightcurves and the
correlation only emerges when averaging many features in a
cross-correlation function.

We show the results in Fig.~\ref{ModelCCFFig}.  Note that both model
CCFs show an oscillatory character.  This is a feature of the X-ray
ACF, and is a consequence of the QPO in the X-ray lightcurves.
Overall, the agreement between model and observed CCFs is good,
supporting the interpretation of the correlation as arising from disc
reprocessing.

\section{Discussion}

We have interpreted correlations seen at the outburst peak as due to
thermal reprocessing and those on the decline as instead associated
with synchrotron emission as inferred for XTE~J1118+480.  These
interpretations are obviously rather speculative and uncertain.  

Thermal reprocessing at the outburst peak is eminently plausible and
the lags are in the range expected give the proposed orbital period of
the binary.  Other X-ray binaries have shown correlations consistent
with reprocessing in the disk (e.g.\
\citet{Hynes:1998a,Hynes:2006a,Hynes:2007a}, and the spectral energy
distribution at outburst peak could be successfully modelled as due to
an accretion disk \citep{Still:2005a}. This can be considered the most
natural interpretation.

Associating the correlations seen on the decline with synchrotron
emission is more tentative.  Primarily we are interpreting the
presence of the dip in the CCF; had this not been present, we would
not have been led to this conclusion.  In XTE~J1118+480, the presence
of optical synchrotron could be associated with a very flat spectral
energy distribution extending into the IR.  This was clearly not the
case early in the outburst of \target\ as the SED was modelled as a
disk.  In fact our SMARTS monitoring data shows no evidence for a
change in the shape of the SED during the period observed, with the
$V-H$ colour remaining approximately constant to within the accuracy of
our measurements.  

The absence of a change in the optical/IR colour does call a
synchrotron interpretation into question.  Given the evolution of the
source between the two epochs there is no reason to expect that the
origin of the correlated variability should be different.
Nonetheless, the CCFs clearly are different very different, and so a
change must have occurred.  Of course, if the decline data are
associated with synchrotron emission, and the SED has not changed, we
cannot rule out synchrotron emission at the peak as well.  This seems
less likely, as argued above, but this possibility cannot be
completely rejected.

\section{Conclusions}

We have presented a study of the correlated X-ray and optical
variability in the black hole candidate \target\ at two epochs, one at
the peak of the outburst (2005 July 6--7) and the other on the decline
phase (2005 August 9).  Data from July shows a smeared and lagged
correlation that can be readily interpreted as arising from thermal
reprocessing in the accretion disk.  Data in August show a dip
and peak structure similar to that seen in XTE~J1118+480.  Much later
in the outburst \target\ showed a strong dip followed by a weak peak
\citep{Durant:2008a} indicating that the CCF morphology varies with
the source state.  If it arises in the same way as in XTE~J1118+480,
we should interpret it as due to optical synchrotron emission possibly
associated with a jet, but in the absence of spectral evidence for a
jet contribution in the optical this remains a somewhat unsatisfying
explanation.  As suggested by \citet{Durant:2008a}, it is possible
that the origin is synchrotron or cyclotron emission from above the
disk rather than from a jet.

\section*{Acknowledgements}

These observations would not have been possible without accommodations
from several directions.  We are grateful to the Argos team for
allowing us to use some of their time on very short notice to obtain
the first epoch of observations.  The second epoch was facilitated
through an award of VLT Director's Discretionary Time.  Finally we are
once again indebted to Jean Swank and the \xte\ team for fitting
the public ToO observation schedule around these serendipitous
observing opportunities.

\label{lastpage}

\end{document}